\documentclass[conference]{IEEEtran}
\IEEEoverridecommandlockouts
\usepackage{cite}
\usepackage{amsmath,amssymb,amsfonts}
\usepackage{xurl}
\usepackage[colorlinks=false, hidelinks]{hyperref}
\usepackage{algorithmic}
\usepackage{graphicx}
\usepackage{textcomp}
\usepackage{xcolor}
\usepackage{booktabs}
\usepackage{comment}
\usepackage{multirow}
\usepackage[utf8]{inputenc}
\usepackage[ruled,vlined]{algorithm2e} % 必须包含这个宏包
\usepackage{amsmath}
\usepackage{amsfonts}
\usepackage{stfloats}
\usepackage{hyperref}
\usepackage[capitalize,noabbrev]{cleveref}
\usepackage{microtype}
\usepackage{pifont}
\newcommand{\cmark}{\ding{51}}
\newcommand{\xmark}{\ding{55}}

\def\BibTeX{{\rm B\kern-.05em{\sc i\kern-.025em b}\kern-.08em
    T\kern-.1667em\lower.7ex\hbox{E}\kern-.125emX}}
\begin{document}

\title{LL-SDR: Low-Latency Speech Enhancement via Discrete Representations\\
}

\author{%
\IEEEauthorblockN{%
Jingyi Li\IEEEauthorrefmark{2},
Luca Della Libera\IEEEauthorrefmark{3}\IEEEauthorrefmark{4},
Mirco Ravanelli\IEEEauthorrefmark{3}\IEEEauthorrefmark{4},
Mingkun Xu\IEEEauthorrefmark{2}\IEEEauthorrefmark{1},
Cem Subakan\IEEEauthorrefmark{4}\IEEEauthorrefmark{5}\IEEEauthorrefmark{1}%
}
\IEEEauthorblockA{\IEEEauthorrefmark{2}Guangdong Institute of Intelligence Science and Technology}
\IEEEauthorblockA{\IEEEauthorrefmark{3}Concordia University}
\IEEEauthorblockA{\IEEEauthorrefmark{4}Mila -- Quebec AI Institute}
\IEEEauthorblockA{\IEEEauthorrefmark{5}Laval University}
\IEEEauthorblockA{\IEEEauthorrefmark{1}Corresponding authors: \{xumingkun@gdiist.cn,\ cem.subakan@ift.ulaval.ca\}}%
}

\maketitle

\begin{abstract}
Many speech enhancement (SE) methods rely on continuous representations. Recently, discrete audio tokens have been explored to enable autoregressive generation for SE. However, it remains unclear whether discretization itself consistently improves SE performance. In this paper, we introduce LL-SDR, a token-based speech enhancement framework that explicitly leverages discretization to better separate speech and noise. Our first contribution is a Variance-Ordered Residual Vector Quantizer (VO-RVQ), designed to disentangle speech and noise distributions during tokenization. Second, we propose a latent-space discriminator to better align enhanced embeddings with semantic embeddings. Experiments show that LL-SDR outperforms continuous baselines and matches the performance of autoregressive token-based approaches. Despite its strong enhancement performance, LL-SDR remains lightweight and efficient, requiring only 40G MACs for a single forward pass on a 10-second 16 kHz speech segment and achieving low-latency inference with an RTF of 0.01 on GPU and 0.24 on CPU. Demos and source code are available at our project websites \footnote{Demos: \url{https://jingyi49.github.io/demo_SE/}.}\footnote{Codes: \url{https://github.com/jingyi49/llsdr}.}.
\end{abstract}

\vspace{-.3cm}
\section{Introduction}
Speech enhancement (SE) aims to improve the quality and intelligibility of speech signals by suppressing noise, reverberation, or other distortions while preserving the underlying speech content. Most existing speech enhancement approaches are built upon continuous acoustic features. These primarily include methods formulated in the time-frequency (T-F) domain~\cite{zhao2024frcrnboostingfeaturerepresentation}, where spectrograms are extracted via the Short-Time Fourier Transform (STFT), as well as those operating directly on raw waveforms~\cite{defossez2019demucsdeepextractormusic, 8707065}. Despite their differences, both categories rely on processing continuous-valued representations.

A novel approach in speech processing is audio tokenization,
which converts audio signals into discrete tokens. Discrete tokens provide a more efficient latent domain than conventional time or time-frequency domain of signals~\cite{yang24h_interspeech}. Inspired by this paradigm, Neural Audio Codec (NAC)-based approaches have recently gained traction in speech enhancement. SELM~\cite{wang2024selmspeechenhancementusing} was the first to incorporate semantic tokens into the speech enhancement pipeline. More recent models, such as LLaSE-G1~\cite{kang2025llaseg1incentivizinggeneralizationcapability} and GenSE~\cite{yao2025gensegenerativespeechenhancement}, formulate speech enhancement as a conditional language modeling task by tokenizing speech into semantic tokens using a pre-trained self-supervised model and acoustic tokens using a neural codec.
By leveraging the powerful contextual modeling capabilities of autoregressive architectures, these models achieve competitive performance in various speech enhancement tasks. This success highlights the significant advantages of discrete representations in capturing complex speech patterns for enhancement tasks.

Despite the superior quality achieved by these LM-based frameworks, the inherent latency of autoregressive architectures remains a critical bottleneck~\cite{sun2020emapproachnonautoregressiveconditional}. Speech enhancement is a foundational technology for a wide range of downstream applications~\cite{defossez2020realtimespeechenhancement}, including telecommunications~\cite{reddy2019scalablenoisyspeechdataset}, hearing aids~\cite{8031044}, and automatic speech recognition (ASR)~\cite{zorila2019investigationeffectivenessenhancementasr}. Since these applications involve human-to-human or human-to-machine interaction, real-time processing with minimal latency is of paramount importance. The autoregressive architecture often fails to meet the low-latency requirement. Li et al. \cite{Li_2025} proposed a novel NAC-based solution designed for real-time speech enhancement using a non-autoregressive architecture. However, rather than operating on discrete tokens, their approach utilizes the continuous embeddings extracted from the pretrained neural codec before the quantization stage, overlooking the improvements that discrete representation could offer. In this study, we provide a comparative analysis of discrete versus continuous representations for speech enhancement within non-autoregressive (NAR) frameworks. Our findings demonstrate that within our proposed non-autoregressive framework, discrete representations consistently outperform continuous baselines in both reverberant and non-reverberant noisy environments.

The main contributions are as follows:
\begin{itemize}
\item \textbf{Low-latency model}: We propose a lightweight, low-latency speech enhancement model that achieves performance comparable to state-of-the-art autoregressive methods such as LLaSE \cite{kang2025llaseg1incentivizinggeneralizationcapability} and GenSE ~\cite{yao2025gensegenerativespeechenhancement}.

\item \textbf{Noise-speech disentanglement through variance-ordered quantization}: We use a progressive masking structure (representations ordered by latent dimensionality) to explicitly enforce an ordering across residual stages. This Variance-Ordered Residual Vector Quantizer (VO-RVQ) can effectively disentangle speech distributions from low-variance noise distributions. This structure demonstrates strong robustness in challenging acoustic scenarios, including reverberant environments and real-world recordings.

\item \textbf{Semantic grounding through a learned discriminator}: To inject semantic information into the discrete space, we introduce a HuBERT-based discriminator for representation alignment. This mechanism ensures that the learned embeddings remain semantically grounded in non-autoregressive frameworks. Compared with earlier approaches this approach is different as it employs Hubert features in the discriminator. 
\end{itemize}

\section{Method}
\subsection{Framework Overview}
\begin{figure*}[t]
  \centering
  \includegraphics[width=0.85\textwidth]{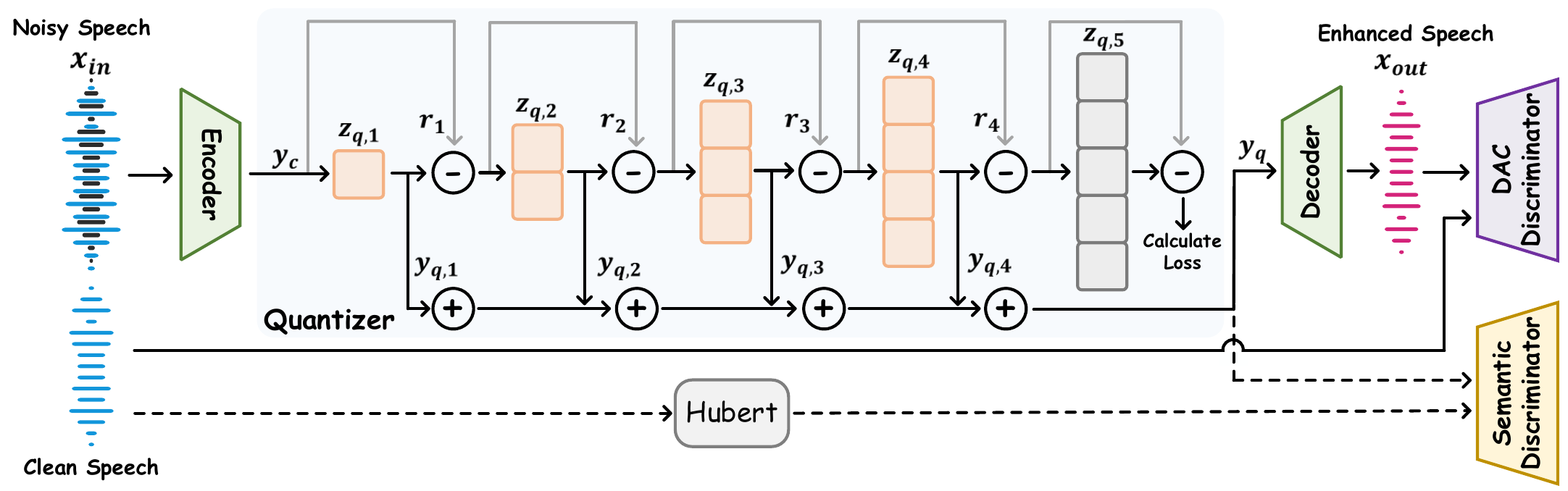}
\vspace{-5pt}
  \caption{LL-SDR speech enhancement framework: a noisy waveform is encoded and enhanced in the discrete space via our variance-ordered quantizer where $N_e$ is four and $N_n$ is one, then decoded back to waveform. The triangle structure here illustrates that different quantizers have different dimensional sizes; a HuBERT-based semantic discriminator to align the enhanced representation with clean speech semantics.}
  \label{fig:architecture}
  \vspace{-10pt}
\end{figure*}
As shown in \cref{fig:architecture}, the generator of our SE architecture consists of a codec encoder, a quantizer, and a decoder. Formally,

\begin{equation}
\label{eq:generator_flow}
\begin{aligned}
    \mathbf{y}_{c} &= \text{DACEncoder}(\mathbf{x}_{in}), \\
\mathbf{y}_{q} &=  \text{RVQ}(\mathbf{y}_{c}), \\
    \mathbf{x}_{out} &= \text{DACDecoder}(\mathbf{y}_{q}),
\end{aligned}
\end{equation}
\noindent where the quantizer is implemented as a residual vector quantizer (RVQ)~\cite{zeghidour2021soundstream} with $N$ sequential codebooks. In RVQ, the continuous latent representation $\mathbf{y}_c$ is approximated sequentially by multiple codebooks, where each stage quantizes the residual error from the previous stage. Specifically, the first $N_e$ codebooks are allocated to enhanced embeddings, while the remaining $N_n$ codebooks model noisy embeddings, with $N_e + N_n = N$.

\subsection{Variance-Ordered Residual Vector Quantizer}
Ordered representation learning~\cite{xu2021anytimesamplingautoregressivemodels,rippel2014learningorderedrepresentationsnested,guo2024socodecsemanticorderedmultistreamspeech} encourages a PCA-like decomposition in which latent dimensions are ordered by variance: early dimensions (high variance) capture the most salient information, while later dimensions (low variance) progressively model the residual. In speech enhancement, we design the quantization process to encourage speech-related information to be captured by the high-variance components, while noise-related variations are more likely to be represented in the residual components.

Following~\cite{xu2021anytimesamplingautoregressivemodels},  the
modified loss function is an order-inducing objective given by
\begin{equation}
\label{eq:ordered_obj}
\begin{aligned}
\mathcal{L}_{\text{ord}} &= \left\lVert \operatorname{Mel}\bigl(\operatorname{DACDecoder}(\mathbf{y}_{q})\bigr) - \operatorname{Mel}(\mathbf{x}) \right\rVert_{2}^{2}\\
&\quad + \sum_{i=1}^N \left\lVert \operatorname{sg}[\hat z_{c,i}] - \hat z_{q,i} \right\rVert_{F}^{2}\\
&\quad + \sum_{i=1}^N \beta \left\lVert \hat z_{c,i} - \operatorname{sg}[\hat z_{q,i}] \right\rVert_{F}^{2}.
\end{aligned}
\end{equation}
\noindent where $\mathbf{x}$ is the clean target speech, $z_{q,i}$ and $z_{c,i}$ are representations that retain only the first $d_i$ dimensions, $\text{sg}[\cdot]$ denotes the stop-gradient operator and $\left\Vert \cdot \right\Vert_{F}$ the Frobenius norm. The first term in \cref{eq:ordered_obj} is a reconstruction loss, which ensures that the final output matches the clean speech target. The second term and third term encourage an ordered representation learning, forcing the first few dimensions to capture the most high-variance information (typically clean speech). The detailed evolution process of the quantized representation $y_q$ is implemented via our proposed Variance-Ordered Residual Vector Quantizer (VO-RVQ), as formally outlined in \cref{alg:vo_rvq}. Specifically, the triangle masking structure means the effective dimensionality increases progressively across codebooks. At each stage, the residual representation is first projected to a shared full-dimensional space. However, instead of treating all latent dimensions equivalently during quantization, we introduce an ordered latent-dimensionality constraint: at stage $i$, only the first $d_i$ latent dimensions are preserved, while the remaining dimensions are masked to zero.

\begin{figure}[t]
  \centering
  \includegraphics[width=0.55\textwidth]{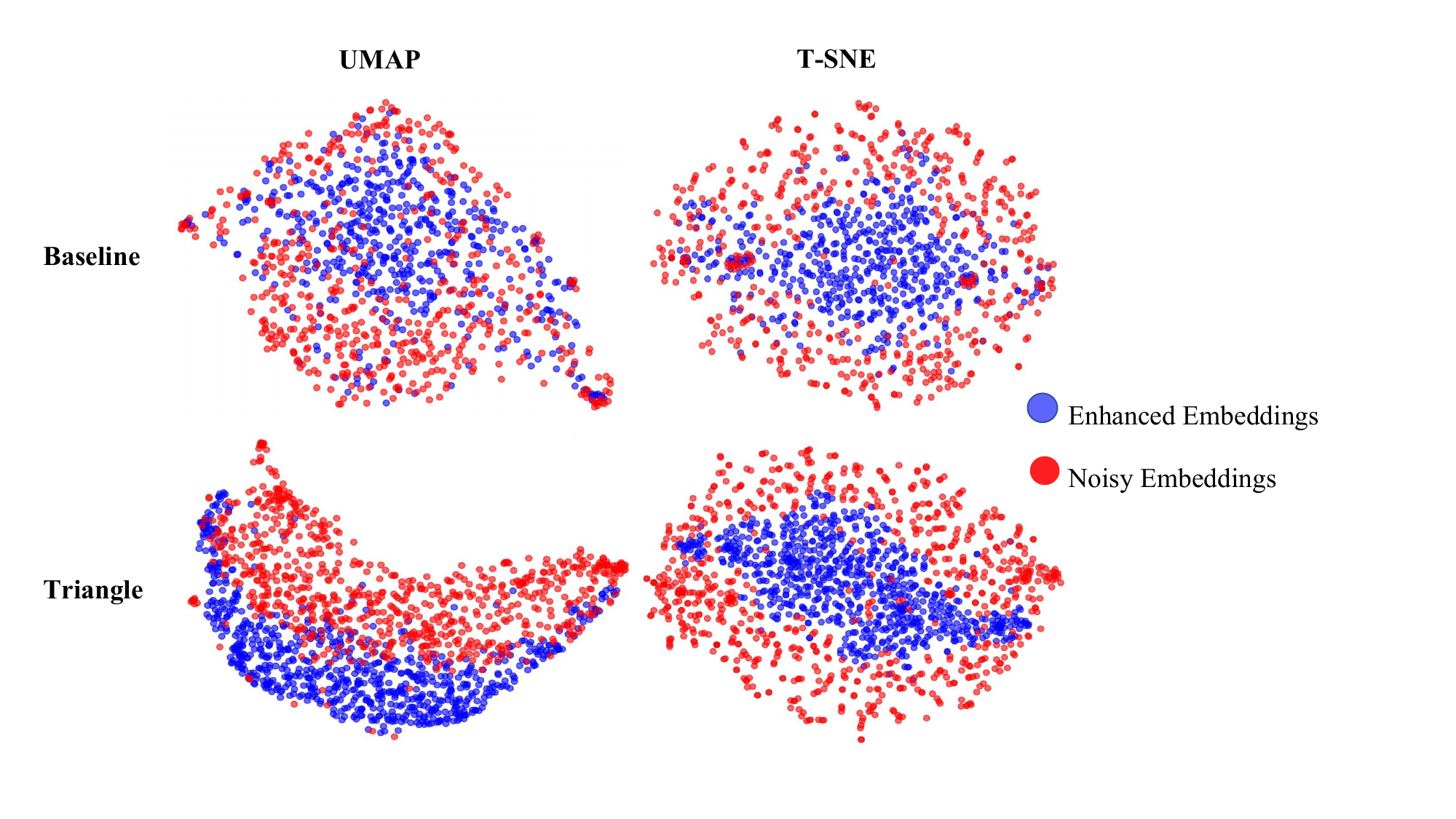}
\vspace{-5pt}
  \caption{Visualization of latent embeddings using UMAP and t-SNE. Blue points denote noisy embeddings (the output of last Nn codebooks), and red points denote enhanced embeddings (the mean of first Ne codebooks). The top row shows the baseline model, where noisy and enhanced representations largely overlap. The bottom row corresponds to the proposed triangle (variance-ordered) quantization structure, which exhibits clearer structural separation and more organized distribution patterns.}
  \label{fig:visualization}
  \vspace{-20pt}
\end{figure}

Since later dimensions are exclusively modeled by higher-index codebooks (larger $i$) in our triangle, these subsequent codebooks (the last $N_n$ codebooks) effectively represent noise components. This ordered representation learning framework facilitates the disentanglement of clean speech from noise components. We further validate this property qualitatively through t-SNE and UMAP visualizations seen in ~\ref{fig:visualization}. In addition, we conduct quantitative spectral clustering experiments in \cref{tab:spectral_clustering}, which further confirm the separability and structured organization of the learned representations.

\SetAlgoSkip{smallskip}
\begin{algorithm}[t]
\SetAlgoLined
\DontPrintSemicolon
\SetKwInOut{Input}{Input}
\SetKwInOut{Output}{Output}
\caption{VO-RVQ}
\label{alg:vo_rvq}

\Input{
    $y_c \in \mathbb{R}^{D_{\text{latent}}}$: Continuous latent; \\
    $\{\mathcal{C}_n\}_{n=1}^{N_e}$: $N_e$ trainable codebooks.
}
\Output{$y_q$: Final quantized representation.}

\BlankLine
$y_q \leftarrow \mathbf{0}_{D_{\text{latent}}}$ \;
$r_0 \leftarrow y_c$ \tcp*{Initialize residual}

\BlankLine
\For{$i = 1$ \KwTo $N$}{
    \BlankLine
    \tcp{Stage $i$: Project and Quantize}
    $z_{c,i} \leftarrow \text{Proj}_{\text{in}}(r_{i-1})$ \hfill ($z_{c,i} \in \mathbb{R}^{d_{\text{full}}}$) \;
    \BlankLine
    \tcp{Triangle masking: Only keep first $d_i$ dims}
    $\hat z_{c,i}$ = $z_{c,i, (1:d_i)}$
    
    \BlankLine
    $\hat{z}_{q,i} \leftarrow \text{Quantize}(\hat z_{c,i}, \mathcal{C}_i)$ \;
    
    \BlankLine
    \tcp{padding to full dimension before projection}
    $z_{q,i}^{\text{padded}} \leftarrow \begin{bmatrix} \hat{z}_{q,i} \\ \mathbf{0} \end{bmatrix}$ \;
    
    \BlankLine
    \tcp{Project back and accumulate enhanced embeddings}
    $y_{q,i} \leftarrow \text{Proj}_{\text{out}}(z_{q,i}^{\text{padded}})$ \;
    if $i \le N_e$, $y_q \leftarrow y_q + y_{q,i}$ \;
    
    \BlankLine
    \tcp{Update residual for next iteration}
    $r_i \leftarrow r_{i-1} - y_{q,i}$ \;
}
\BlankLine
\Return $y_q$
\end{algorithm}

\subsection{Semantic Discriminator}

Following the idea of incorporating semantic supervision from self-supervised speech models, we adopt HuBERT \footnote{https://huggingface.co/facebook/hubert-base-ls960} ~\cite{hsu2021hubertselfsupervisedspeechrepresentation} as a semantic teacher, similar to SpeechTokenizer~\cite{zhang2024speechtokenizerunifiedspeechtokenizer}. However, instead of injecting semantic supervision directly inside the tokenizer, we introduce a \emph{semantic discriminator} that aligns the enhanced representations with HuBERT features.
Specifically, we employ two complementary objectives: (i) an $\ell_1$ regression loss to align feature representations and (ii) an InfoNCE contrastive loss to preserve discriminative semantic structure.

As in SpeechTokenizer, we first enforce feature-level alignment between the enhanced embeddings and the semantic representations extracted by the HuBERT teacher. The alignment loss is defined as
\begin{equation}
\label{eq:l1_proj_align}
\mathcal{L}_{\text{sem}}
=
\left\lVert 
\mathrm{proj}(\mathbf{y_q}) - \text{HuBERT}(\mathbf{x})
\right\rVert_1,
\end{equation}
\noindent where $\mathbf{y_q}$ denotes the enhanced embedding produced by our model. The projection layer $\mathrm{proj}(\cdot)$ is a learnable linear mapping that projects $\mathbf{y_q}$ into the semantic feature space. The projection output is constrained to dimension 768 to match the hidden representation size of the pre-trained HuBERT model.

While $\mathcal{L}_{\text{sem}}$ encourages overall feature similarity, regression-based objectives may lead to over-smoothed representations~\cite{ren2022revisitingoversmoothnesstextspeech}. To preserve discriminative semantic structure across feature dimensions, we additionally introduce a contrastive objective based on Noise Contrastive Estimation (NCE)~\cite{ma2018noisecontrastiveestimationnegative}.

We treat each time step as an independent training sample and flatten the batch and temporal dimensions, resulting in $K = B \times T$ feature vectors. After applying $\ell_2$ normalization to stabilize training and convert dot products into cosine similarities, we define
\begin{equation}
f_{\text{fake}} = \mathrm{norm}(\mathrm{proj}(\tilde{\mathbf{y}}_q)), 
\quad
f_{\text{real}} = \mathrm{norm}(\text{HuBERT}(\tilde{\mathbf{x}})),
\end{equation}
\noindent where $\mathrm{norm}(\cdot)$ denotes $\ell_2$ normalization along the feature dimension.
The contrastive objective is then defined as
\begin{equation}
\label{eq:infonce}
\mathcal{L}_{\mathrm{NCE}}
=
- \mathbb{E}_{i}
\left[
\log
\frac{
\exp(s_{ii})
}{
\sum_{j} \exp(s_{ij})
}
\right],
\quad
s_{ij} = \frac{f_{\text{fake}}^{(i)} \cdot f_{\text{real}}^{(j)}}{\tau},
\end{equation}
\noindent where $\tau$ denotes a temperature parameter controlling the sharpness of the similarity distribution.

\subsection{Training Details}
We follow the training recipe of DAC~\cite{kumar2023highfidelityaudiocompressionimproved} for both data preprocessing and optimization. Notably, all reconstruction terms are defined with respect to the clean target speech rather than the noisy input. The overall objective is a weighted sum of multiple terms, including waveform reconstruction, multi-resolution spectral losses, adversarial losses, and the proposed representation-alignment losses:
\begin{equation}
\label{eq:total_loss}
\mathcal{L}_{\text{total}}
=
\lambda_{\text{ord}}\mathcal{L}_{\text{ord}}
\;+\;
\lambda_{\text{adv}}\mathcal{L}_{\text{adv}}
\;+\;
\lambda_{\text{align}}\left(\mathcal{L}_{\text{NCE}} + \alpha \mathcal{L}_{\text{sem}}\right),
\end{equation}
\noindent where $\lambda_{\text{ord}}$, $\lambda_{\text{adv}}$, and $\lambda_{\text{align}}$ are scalar weights, and $\alpha$ controls the relative strength of the $\ell_1$ alignment term. 
We set the ratio between $N_e$ and $N_n$ according to the empirical variance ratio between speech and noise estimated from the training set. We further discuss the impact of this ratio through experiments in Figure \ref{fig:ratio_notebooks}. 
%As illustrated in \cref{fig:architecture}, four codebooks are allocated to model speech components and one codebook is allocated to model noise, resulting in a 4:1 allocation ratio.
\section{Experiments}

\begin{table*}[t]
\centering
\caption{DNSMOS (OVRL / SIG / BAK) under reverberant, non-reverberant, and real-recording conditions. Most values for compared methods are taken from prior work~\cite{kang2025llaseg1incentivizinggeneralizationcapability} and ~\cite{li2024masksrmaskedlanguagemodel}. Values marked with * are obtained from our own evaluation.
\textbf{Best} and \underline{second-best} results are highlighted.}
\label{tab:mos_main}
\vspace{-5pt}

\setlength{\tabcolsep}{7pt}

%\resizebox{0.70\linewidth}{!}{
\begin{tabular}{lcc ccc ccclll}
\toprule
\multirow{2}{*}{Model} & \multirow{2}{*}{Type} & \multirow{2}{*}{Discrete}
& \multicolumn{3}{c}{Reverb} & \multicolumn{3}{c}{No Reverb} & \multicolumn{3}{c}{Real}\\
\cmidrule(lr){4-6} \cmidrule(lr){7-9} \cmidrule(lr){10-12}
 & & & OVRL $\uparrow$ & SIG $\uparrow$ & BAK $\uparrow$  & OVRL $\uparrow$ & SIG $\uparrow$ & BAK $\uparrow$  & OVRL $\uparrow$ & SIG $\uparrow$ & BAK $\uparrow$\\
\midrule

Noisy        & --  & -- & 1.39 & 1.76 & 1.50  & 2.48 & 3.39 & 2.62  & 2.26
& 3.05
&2.51
\\

\midrule
Conv-TasNet  & NAR & \xmark  & 2.01 & 2.42 & 2.71  & 3.00 & 3.09 & 3.34  & -& -&-\\
DEMUCS       & NAR & \xmark  & 2.55 & 2.86 & 3.90  & 3.35 & 3.58 & 4.15  & 2.99& 3.26
&\underline{4.03}
\\
FRCRN        & NAR & \xmark  & 2.28 & 2.93 & 2.92  & 3.34 & 3.58 & 4.13  & 3.04
& 3.37
&3.98
\\
 VoiceFixer& NAR& \xmark& 3.13& 3.43&  4.02& 3.25& 3.50&4.11 & 3.08*& 3.37*&4.00*\\

%\midrule
%SGMSE        & Diffusion & \xmark  &  2.43 & 2.73 & 2.74  & 3.14 & 3.50 & 3.71  &  2.79& 3.30&2.89\\
%StoRM       & Diffusion & \xmark  & 2.52 & 2.95 & 3.14  & 3.21 & 3.51 & 3.94  &  2.94& 3.41& 3.38\\

\midrule
SELM         & AR  & \cmark  & 2.70 & 3.16 & 3.58  & 3.26 & 3.51 & 4.10  & 3.12& \textbf{3.59}&3.44
\\
MaskSR       & AR  & \cmark  & 3.25 & 3.53 & 4.07  & 3.34 & 3.59 & 4.12  & \underline{3.14}
& 3.43
&\underline{4.03}
\\
AnyEnhance   & AR  & \cmark  & 3.20 & 3.50 & 4.04  & \underline{3.42} & 3.64 & \textbf{4.18}  & -& -&-\\
GenSE        & AR  & \cmark  & 3.19 & 3.49 & 3.73  & \textbf{3.43} & \underline{3.65} & \textbf{4.18}  & -& -&-\\
LLaSE-G1     & AR  & \cmark  & \textbf{3.33} & \textbf{3.59} & \textbf{4.10}  & \underline{3.42} & \textbf{3.66} & \underline{4.17}  & 3.06*& 3.36*&3.96*\\

\midrule
\textbf{LL-SDR (Ours)} & NAR & \cmark & \underline{3.27} & \underline{3.55} & \underline{4.08}  & 3.39 & 3.62 & 4.16  & \textbf{3.16}
& \underline{3.45}
&\textbf{4.05}
\\

\bottomrule
\end{tabular}
%}
\vspace{-10pt}
\end{table*}

\begin{table}[t]
\centering
\caption{Model complexity and runtime comparison. \textbf{Best} and \underline{second-best} results are highlighted.}
\vspace{-5pt}
\label{tab:efficiency}
\setlength{\tabcolsep}{2pt}

\resizebox{\linewidth}{!}{
\begin{tabular}{lcccc}
\toprule
Model 
& Params (M) $\downarrow$
& MACs (G) $\downarrow$
& RTF (H100) $\downarrow$
& RTF (CPU) $\downarrow$ \\
\midrule

FRCRN
& \textbf{53}& \underline{77}& 0.04 
& 0.99\\

VoiceFixer 
& 117& 777& \textbf{0.01}& \underline{0.52}\\

LLaSE-G1 (single)
& 1319
& 145 
& 0.37
& 2.43\\

\midrule
\textbf{LL-SDR (Ours)}
& \underline{74}& \textbf{40}& \textbf{0.01}
& \textbf{0.24} \\

\bottomrule
\end{tabular}
}

\end{table}

% Requires: \usepackage{multirow}

\begin{table}[t]
\centering
\caption{In-distribution Test:Distortion Metrics Under Reverberant and Non-Reverberant Conditions in DNS-Challenge 2020 Non-Blind Test Set.}
\vspace{-5pt}
\label{tab:distortion_metrics}
\setlength{\tabcolsep}{2pt}

\resizebox{\linewidth}{!}{
\begin{tabular}{lcccccc}
\toprule
\multirow{2}{*}{Model} & \multicolumn{3}{c}{Reverb} & \multicolumn{3}{c}{Non-reverb} \\
\cmidrule(lr){2-4} \cmidrule(lr){5-7}
& STFT $\downarrow$ & Mel$\downarrow$ & LPS$\downarrow$ & STFT$\downarrow$ & Mel$\downarrow$ & LPS$\downarrow$ \\
\midrule
Noisy   & 244.41 & 90.78 & 20.43 & 182.97 & 73.15 & 13.97 \\
FRCRN   & 216.18 & 76.37 & 13.72 & \textbf{125.17} & \textbf{44.39} & \textbf{3.60}  \\
 VoiceFixer & \textbf{192.51}& \textbf{68.98}&\textbf{9.77}& 180.40&63.22 &8.24\\
LLaSE-G1   & 258.73 & 99.10 & 20.32 & 216.47 & 83.54 & 12.79 \\
\midrule
\textbf{LL-SDR (Ours)}  & \underline{208.44}& \underline{74.47}& \underline{11.85}& \underline{161.95}& \underline{53.26}& \underline{6.40}\\ 
\bottomrule
\end{tabular}
}
\end{table}

% Requires: \usepackage{booktabs}

%cem{add rtf}
\begin{table}[h]
\centering
%\caption[Out-of-distribution distortion metrics on the VoiceBank-DEMAND test set]{Out-of-distribution distortion metrics on the VoiceBank-DEMAND test set.}
\caption{Out-of-distribution distortion metrics on the VoiceBank-DEMAND test set.}
\label{tab:voicebank}
\resizebox{0.9\columnwidth}{!}{
\begin{tabular}{lllccc}
    \toprule
    Method &   RTF (H100) $\downarrow$&RTF (CPU) $\downarrow$&STFT $\downarrow$ & Mel $\downarrow$ & LPS $\downarrow$ \\
    \midrule
    Noisy      &   -&-&155.72 & 54.23 & 14.76 \\
    FRCRN      &   0.04 &0.99&\textbf{100.10} & \textbf{29.81} & \textbf{4.93}  \\
    VoiceFixer &   \textbf{0.01}&\underline{0.52}&\underline{123.53} & \underline{30.86} & \underline{6.56}  \\
    LLaSE      &   0.37&2.43&183.25 & 60.55 & 13.94 \\
    \midrule
    LL--SDR    &   \textbf{0.01}&\textbf{0.24}&136.84 & 36.77 & 8.15  \\
    \bottomrule
\end{tabular}
}
\vspace{2pt}

\footnotesize
\emph{Note:} The test set is from the VoiceBank-DEMAND-16k dataset: \url{https://huggingface.co/datasets/JacobLinCool/VoiceBank-DEMAND-16k}.
\end{table}
As shown in Tables~\ref{tab:distortion_metrics} and~\ref{tab:voicebank}, LL-SDR consistently achieves lower distortion than the noisy input and the generative baseline LLaSE-G1 on both in-distribution and out-of-distribution test sets, demonstrating its effectiveness in reducing spectral distortion while maintaining competitive enhancement quality. Qualitative results (see demo page) further suggest that our method better preserves speaker characteristics compared to LLaSE, leading to higher perceived speaker similarity and more faithful reconstruction of the source voice. 

\begin{table}[t]
\centering
\caption{Spectral clustering results. Enhanced and noisy embeddings are first combined without labels, clustered using spectral clustering, and then compared with the true labels.}
\label{tab:spectral_clustering}
\vspace{-5pt}
\setlength{\tabcolsep}{5pt}

\resizebox{\linewidth}{!}{
\begin{tabular}{lccc}
\toprule
Method & Accuracy $\uparrow$ & Macro Recall $\uparrow$ & Macro F1 $\uparrow$ \\
\midrule
RVQ & 50.46 & 50.46 & 34.37 \\
VO-RVQ & \textbf{71.33} & \textbf{71.33} & \textbf{71.25} \\
\bottomrule
\end{tabular}
}
\vspace{-10pt}
\end{table}

\begin{table}[t]
\centering
\caption{Ablation study One. \textbf{Best} and \underline{second-best} results are highlighted.}
\label{tab:ablation_mos}
\vspace{-5pt}
\setlength{\tabcolsep}{2pt}

\resizebox{\linewidth}{!}{
\begin{tabular}{l ccc ccc}
\toprule
\multirow{2}{*}{Method}
& \multicolumn{3}{c}{Reverb}
& \multicolumn{3}{c}{No Reverb} \\
\cmidrule(lr){2-4} \cmidrule(lr){5-7}
& OVRL $\uparrow$ & SIG $\uparrow$ & BAK $\uparrow$ & OVRL $\uparrow$ & SIG $\uparrow$ & BAK $\uparrow$ \\
\midrule

Continuous & 2.95 & 3.30 & 3.84 & 3.30 & 3.56 & 4.09 \\
RVQ      & 3.13 & 3.43 & \underline{4.01} & 3.36 & 3.59 & \underline{4.16} \\
VO-RVQ   & \underline{3.14} & \underline{3.45} & {3.99} & \underline{3.39} & \underline{3.62} & \textbf{4.17} \\

\midrule
VO-RVQ + HuBERT
& \textbf{3.22} & \textbf{3.51} & \textbf{4.03}
& \textbf{3.40} & \textbf{3.63} & \underline{4.16} \\

\bottomrule
\end{tabular}
}
\vspace{-10pt}
\end{table}

\begin{table}[t]
\centering
\caption{Ablation study Two. \textbf{Best} results are highlighted.}
\label{tab:ablation_loss}
\vspace{-5pt}
\setlength{\tabcolsep}{2pt}

\resizebox{\linewidth}{!}{
\begin{tabular}{l ccc ccc}
\toprule
\multirow{2}{*}{Method}
& \multicolumn{3}{c}{Reverb}
& \multicolumn{3}{c}{No Reverb} \\
\cmidrule(lr){2-4} \cmidrule(lr){5-7}
& OVRL $\uparrow$ & SIG $\uparrow$ & BAK $\uparrow$ & OVRL $\uparrow$ & SIG $\uparrow$ & BAK $\uparrow$ \\
\midrule

w/o $\mathcal{L}_{\text{NCE}}$ & 3.15 & 3.45 & 4.01 & 3.37 & 3.61 & 4.15\\
w/o $\mathcal{L}_{\text{sem}}$ & 3.13 & 3.45 & 3.96 & 3.38 & 3.61 & 4.16\\
\midrule
\textbf{LL-SDR (Ours)}& \textbf{3.27}& \textbf{3.55}& \textbf{4.08}& \textbf{3.39}& \textbf{3.62}& \textbf{4.16}\\

\bottomrule
\end{tabular}
}
\vspace{-10pt}
\end{table}

\begin{table}[t]
\centering
\caption{Ablation study Three. \textbf{Best} results are highlighted.}
\label{tab:ablation_shuffle}
\vspace{-5pt}
\setlength{\tabcolsep}{2pt}

\resizebox{\linewidth}{!}{
\begin{tabular}{l ccc ccclll}
\toprule
\multirow{2}{*}{Method}
& \multicolumn{3}{c}{Reverb}
& \multicolumn{3}{c}{No Reverb} &  \multicolumn{3}{c}{Real}\\
\cmidrule(lr){2-4} \cmidrule(lr){5-7} \cmidrule(lr){8-10}
& OVRL $\uparrow$ & SIG $\uparrow$ & BAK $\uparrow$ & OVRL $\uparrow$ & SIG $\uparrow$ & BAK $\uparrow$  & OVRL $\uparrow$ & SIG $\uparrow$ &BAK $\uparrow$  \\
\midrule

Shuffle One& 3.19& 3.50& 4.00& 3.38& 3.61& 4.15 & 3.12& 3.44&3.99\\
 Shuffle Two& 3.12& 3.44& 3.97& 3.38
& 3.62&4.15 & 3.12& 3.43&4.00\\
 Shuffle Three& 3.18& 3.48& 4.00& 3.36& 3.60&4.14 & 3.12& 3.44&3.99\\

\midrule
\textbf{LL-SDR (Ours)}& \textbf{3.27}& \textbf{3.55}& \textbf{4.08}& \textbf{3.39}& \textbf{3.62}& \textbf{4.16} & \textbf{3.16}& \textbf{3.45}&\textbf{4.05}\\

\bottomrule
\end{tabular}
}
\vspace{-10pt}
\end{table}

%more robust decomposition/more structured latent representation, suitable for more challenging data
\subsection{Dataset}
For the ablation study in \cref{tab:ablation_mos}, we train our model on LibriSpeech-100~\cite{7178964} and the DNS Challenge~\cite{reddy2020interspeech} datasets. Room impulse responses (RIRs) are sourced from the DNS Challenge dataset, and noise signals are also drawn from DNS. For comparisons against other methods, we additionally train a larger variant using LibriSpeech-360 together with DNS; in this setting, noise is sampled from both DNS and DEMAND~\cite{thiemann2013demand}, while RIRs remain from DNS. For evaluation, all evaluation sets are taken from the DNS Challenge 2020. We set the ratio of reverberant to non-reverberant data in the training sets to 1:1. Training segments are randomly cropped to 0.38 s, whereas validation segments use durations of 5.0 s, respectively. 

\subsection{Metrics}
We evaluate speech quality using DNSMOS~\cite{reddy2021dnsmosnonintrusiveperceptualobjective}. The evaluation also employs three distortion metrics to assess the quality of enhanced speech. \textbf{STFT} Distance measures the Euclidean distance between the Short-Time Fourier Transform (STFT) spectrograms of the enhanced and clean signals, capturing discrepancies in the time-frequency domain. \textbf{Mel} Distance computes the distance in the mel-scale spectrogram space, which aligns more closely with human auditory perception by emphasizing lower frequencies. \textbf{LPS} (Log Power Spectrum) Distance quantifies the distortion in the logarithmic power spectrum domain, reflecting energy distribution differences across frequency bins. Lower values across all three metrics indicate better speech enhancement performance. 

We do not compute sample-level metrics in time-domain such as PESQ, SI-SNR, and STOI, as these metrics are not well suited for generative methods. Sample-level metrics such as PESQ evaluates quality by directly comparing the enhanced speech to its clean counterpart in time domain, measuring sample-level similarity. However, generative models aim to model the distribution of real data, and as a result, the generated speech might sound realistic despite having a more considerable distance from the clean version ~\cite{yao2025gensegenerativespeechenhancement,kang2025llaseg1incentivizinggeneralizationcapability}. In generative settings, the synthesized waveform may not be perfectly aligned with the reference signal even when the perceptual quality is high, often resulting in artificially low scores~\cite{wang2024selmspeechenhancementusing,libera2025focalcodec,dellalibera2025focalcodecstreamstreaminglowbitratespeech}. Therefore, we do not report these metrics in our evaluation.

%However, generative models aim to model the distribution of real data, and as a result, the generated speech might sound realistic despite having a more considerable distance from the clean version.

%Sample-level metrics such as PESQ don’t use in this paper: The key issue lies in the fact that PESQ evaluates quality by directly comparing the enhanced speech to its clean counterpart, measuring sample-level similarity. However, generative models aim to model the distribution of real data, and as a result, the generated speech might sound realistic despite having a more considerable distance from the clean version.[1] Therefore, intrusive metrics such as PESQ and STOI may not provide a precise evaluation of signal quality for generative models. [2,3,4,5]

\subsection{Baselines}
We compare our method with representative continuous SE baselines, including Conv-TasNet~\cite{8707065}, Demucs~\cite{defossez2019demucsdeepextractormusic},  FRCRN~\cite{zhao2024frcrnboostingfeaturerepresentation}, and VoiceFixer ~\cite{Liu_2022}, as well as token-based SE systems that operate on discrete tokens, including SELM~\cite{wang2024selmspeechenhancementusing}, MaskSR~\cite{li2024masksrmaskedlanguagemodel}, AnyEnhance~\cite{Zhang_2025}, GenSE~\cite{yao2025gensegenerativespeechenhancement}, and LLaSE-G1~\cite{kang2025llaseg1incentivizinggeneralizationcapability}.

\subsection{Results and Discussion}
As shown in \cref{tab:mos_main}, our method significantly outperforms conventional non-autoregressive enhancement models such as Conv-TasNet and FRCRN across all DNSMOS metrics. In particular, LL-SDR achieves strong improvements in overall perceptual quality (OVRL) and signal quality (SIG) under both reverberant and non-reverberant conditions. 
Despite being a non-autoregressive model, our approach achieves performance comparable to recent autoregressive generative methods such as GenSE and LLaSE-G1. Notably, LL-SDR attains the second-best results under reverberant conditions and the best results in the real-recordings setting, demonstrating its robustness and strong  capability in complex environments. As shown in Table~\ref{tab:urgent_challenge}, LL-SDR achieves the best OVRL score while maintaining competitive NISQA and UTMOS scores.

\cref{tab:efficiency} compares the computational efficiency of the evaluated models. Real-time factor (RTF) is measured on identical 10-second speech segments with a batch size of 1, using both a single NVIDIA H100 GPU and a CPU platform equipped with an AMD EPYC 9454 processor with 48 cores across 4 NUMA nodes. Each measurement is repeated 5 times, and the average RTF is reported. Our model achieves the lowest CPU RTF, substantially faster than both conventional enhancement models and autoregressive approaches. On GPU, LL-SDR remains competitive, same as VoiceFixer and significantly faster than FRCRN and LLaSE-G1.

% Requires: \usepackage{booktabs}
\begin{table}[h]
    \centering
    \caption{Speech enhancement results on the 2025 URGENT Challenge blind test set. }
    \label{tab:urgent_challenge}
    \begin{tabular}{lcccc}
        \toprule
        Team/Model & Team Rank & OVRL & NISQA & UTMOS \\
        \midrule
        Bobbsun    & 1  & 2.88 & 3.22 & \textbf{2.09} \\
        rc         & 2  & 2.83 & 2.92 & 2.03  \\
        \midrule
        FRCRN      & NA & 1.52 & 1.30 & 1.32  \\
        VoiceFixer & NA & 2.83 & \textbf{3.40} & 1.98  \\
        LLaSE--G1  & NA & 2.62 & 3.02 & 1.94  \\
        \midrule
       \textbf{LL-SDR (Ours)}   & NA & \textbf{2.93} & 2.92 & 1.95  \\
        \bottomrule
    \end{tabular}
    
    \vspace{2pt}
    \footnotesize
    \emph{Note:} The ranks are taken from the leaderboard, which takes into account both non-intrusive and intrusive metrics. Bobbsun and rc are the top two participating teams in the URGENT Challenge leaderboard. RTFs are not reported because the model weights of these leaderboard systems are not publicly available.
\end{table}
%\cem{We need to mention that we are comparing with the top two participants to the challenge and we do not have the weights to calculate the RTF.}

Model complexity is reported in terms of parameters and multiply-accumulate operations (MACs). MACs are computed using \texttt{ptflops}\footnote{\href{https://pypi.org/project/ptflops}{https://pypi.org/project/ptflops}}. LL-SDR requires more parameters than FRCRN, it remains considerably lighter than LLaSE-G1 and VoiceFixer while offering comparable perceptual quality. Overall, these results highlight that the proposed approach achieves a favorable trade-off between SE performance and efficiency.

\subsection{Disentanglement Analysis}
\label{exp:disentaglement}
To evaluate whether the proposed representation disentangles clean speech and noise, we perform a clustering analysis on the learned embeddings. For each sample, we average the embeddings from the first $N_e$ codebooks to obtain the enhanced representation, $\frac{1}{N_e}\sum_{i=0}^{N_e-1} z_{q,i}^{\text{padded}}$, and average the embeddings from the remaining $N_n$ codebooks to obtain the noise representation, $\frac{1}{N_n}\sum_{i=N_e}^{N-1} z_{q,i}^{\text{padded}}$. These two sets of embeddings are then combined into a balanced 1:1 set and clustered using spectral clustering without access to the ground-truth labels. Spectral clustering is performed solely based on the representation geometry, and the resulting cluster assignments are finally compared with the clean/noise labels. A high clustering accuracy indicates that the enhanced and noise embeddings form separable groups in the learned representation space, suggesting that the proposed representation effectively disentangles clean speech and noise.

%Although the first N_e and remaining N_n codebooks are assigned to enhanced and noisy embeddings, respectively, this partition is not used in the clustering stage. For each sample, we average embeddings within the N_e and N_n groups to form a balanced 1:1 set, then mix them and perform spectral clustering without access to codebook indices. The clustering thus relies purely on representation geometry. If the separation were merely induced by the predefined partition, such unsupervised clustering would not reliably recover the groups after mixing. Therefore, the observed separability—also consistent with UMAP visualization in the demo page supports meaningful disentanglement. 

As shown in \cref{tab:spectral_clustering}, the baseline RVQ representation achieves near-random performance (50.46\% accuracy), indicating that clean and noisy embeddings are not well separated in the latent space. In contrast, the proposed VO-RVQ achieves 71.33\% accuracy and a Macro F1 score of 71.25\%, demonstrating substantially improved separability.
This improvement suggests that the triangle-shaped quantization structure encourages the latent space to organize clean speech and noise components into more distinct clusters, supporting the disentanglement hypothesis. In particular, the variance ordering promotes early dimensions to capture high-variance speech-dominant structure, while later dimensions progressively encode residual noise components.

\subsection{Ablation Study}
\label{exp:ablation}
We conduct an ablation study to quantify the contribution of discretization, variance-ordered quantization, and semantic alignment. Starting from a continuous baseline, we progressively introduce RVQ, the proposed VO-RVQ, and HuBERT-based semantic guidance.

As shown in \cref{tab:ablation_mos}, discretization already provides consistent improvements over the continuous baseline across all DNSMOS metrics under both reverberant and non-reverberant conditions. Replacing standard RVQ with the proposed VO-RVQ further improves overall and signal quality scores, suggesting that variance ordering leads to more informative and structured latent representations. Finally, incorporating HuBERT-based semantic alignment yields the best overall performance, achieving the highest scores in most metrics.

%\cem{We should add a note that $L_NCE$ helps, which shows that our novel idea contributes to the performance}
\begin{figure}[t]
  \centering
  \includegraphics[width=0.5\textwidth]{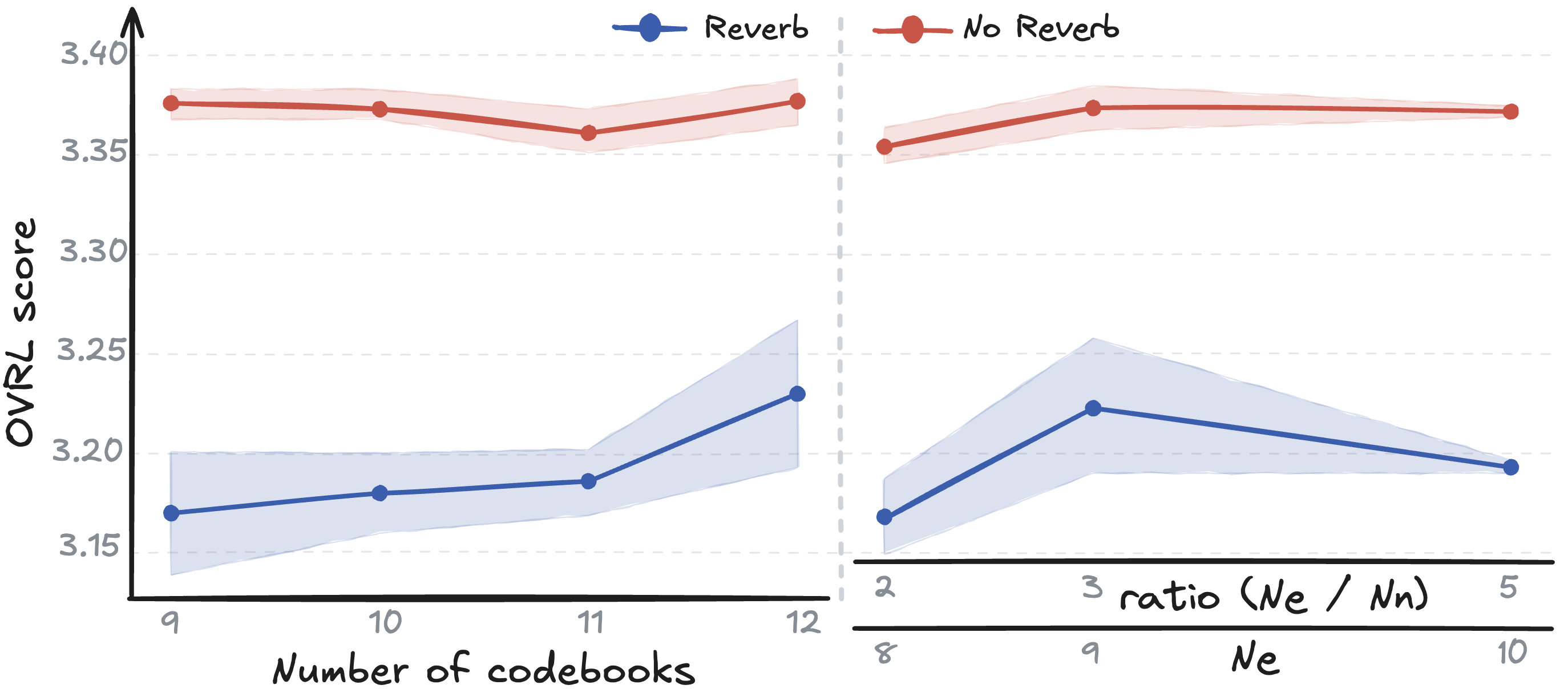}
\vspace{-5pt}
  \caption{Left: DNSMOS OVRL scores for different Number of codebooks settings. Right: DNSMOS OVRL scores for different $N_e$ settings.OVRL scores are averaged over three independent runs for each setting. The solid lines indicate the mean OVRL scores under reverberant and non-reverberant conditions, while the shaded regions represent the standard deviation across the three runs.}
  \label{fig:ratio_notebooks}
  \vspace{-20pt}
\end{figure}

As shown in Table~\ref{tab:ablation_loss}, removing $\mathcal{L}_{\mathrm{sem}}$ leads to a degradation in both reverberant and non-reverberant conditions, indicating that the semantic alignment loss is important for preserving speech-related information. Removing the proposed $\mathcal{L}_{\mathrm{NCE}}$ also consistently degrades the performance, especially under reverberant conditions, where the OVRL score drops from 3.27 to 3.15 and the BAK score drops from 4.08 to 4.01. 
These results demonstrate that the proposed contrastive objective provides complementary benefits beyond the standard L1-based semantic alignment loss. By jointly optimizing $\mathcal{L}_{\mathrm{sem}}$ and $\mathcal{L}_{\mathrm{NCE}}$, LL-SDR achieves the best overall performance, confirming the effectiveness of the proposed training objective.

To further investigate whether the Variance-Ordered Residual
Vector Quantizer (VO-RVQ) -- the ordered structure of latent dimensionality -- is important, we conduct an ablation study by manually shuffling the order of the learned dimensions. Specifically, the original LL-SDR keeps the ordered representation as [1, 2, 3, 4, 5, 6, 7, 8, 9, 10, 11, 12], while the shuffled variants randomly permute the dimension order, such as [1, 11, 10, 9, 8, 6, 7, 5, 4, 3, 2, 12] for Shuffle One and [1, 10, 11, 9, 8, 6, 5, 7, 4, 2, 3, 12] for Shuffle Two/Three.

As shown in Table~\ref{tab:ablation_shuffle}, disrupting the ordered latent dimensions consistently leads to significant performance degradation under both reverberant and real-recording conditions. In contrast, LL-SDR achieves the best results across all metrics when the original ordered representation is preserved. These results indicate that the dimension order learned by our method is not arbitrary, but instead encodes meaningful structural information for speech enhancement, . This is particularly important in challenging acoustic scenarios, especially in complicated environments such as reverberant environments and real-world recordings, demonstrating the robustness of our dimensionality-ordered representation learning strategy.

To analyze the impact of the key configuration parameters, we visualize the DNSMOS scores under both reverberant and non-reverberant conditions. The figure~\ref{fig:ratio_notebooks} compares different Ratio $\frac{N_e}{N_n}$ settings, where the ratio from 2 to 3 improves all three DNSMOS metrics under both conditions, while further increasing it to 5 does not bring additional gains. The best overall performance is obtained at Ratio 3, achieving the best performance. This ratio matches the variance ratio between speech and noise estimated from the training dataset we used. The figure~\ref{fig:ratio_notebooks} also studies the effect of the number of codebooks. As the number of codebooks increases from 9 to 12, the DNSMOS scores generally improve, especially under reverberant conditions.

%\begin{figure}[t]
%  \centering
%  \includegraphics[width=0.45\textwidth]{ratio_2.pdf}
%\vspace{-5pt}
%  \caption{DNSMOS OVRL scores for different $N_e$ settings.}
%  \label{fig:ratio_impact}
%  \vspace{-20pt}
%\end{figure}

\section{Conclusion}
In this paper, we introduced LL-SDR, a low-latency non-autoregressive speech enhancement model that achieves performance comparable to autoregressive approaches. Our method employs a variance-ordered quantizer that induces ordered latent dimensions, encouraging speech and noise components to be disentangled in the latent space, prioritizing high-information speech representations while suppressing noise patterns. To preserve semantic information, we incorporate a HuBERT-based semantic discriminator that aligns the enhanced representations with semantic embeddings from clean speech. Experiments show that the proposed approach achieves competitive perceptual quality under both reverberant, non-reverberant and real-recordings conditions while significantly improving inference efficiency compared to autoregressive baselines. These results demonstrate the potential of structured discrete representations for efficient speech enhancement in complex environments.

%\section*{References}
\begingroup
\sloppy
\bibliographystyle{IEEEtran}
\bibliography{references}

\begin{thebibliography}{10}
\providecommand{\url}[1]{#1}
\csname url@samestyle\endcsname
\providecommand{\newblock}{\relax}
\providecommand{\bibinfo}[2]{#2}
\providecommand{\BIBentrySTDinterwordspacing}{\spaceskip=0pt\relax}
\providecommand{\BIBentryALTinterwordstretchfactor}{4}
\providecommand{\BIBentryALTinterwordspacing}{\spaceskip=\fontdimen2\font plus
\BIBentryALTinterwordstretchfactor\fontdimen3\font minus \fontdimen4\font\relax}
\providecommand{\BIBforeignlanguage}[2]{{%
\expandafter\ifx\csname l@#1\endcsname\relax
\typeout{** WARNING: IEEEtran.bst: No hyphenation pattern has been}%
\typeout{** loaded for the language `#1'. Using the pattern for}%
\typeout{** the default language instead.}%
\else
\language=\csname l@#1\endcsname
\fi
#2}}
\providecommand{\BIBdecl}{\relax}
\BIBdecl

\bibitem{zhao2024frcrnboostingfeaturerepresentation}
S.~Zhao, B.~Ma, K.~N. Watcharasupat, and W.~S. Gan, ``{FRCRN}: Boosting feature representation using frequency recurrence for monaural speech enhancement,'' \emph{IEEE International Conference on Acoustics, Speech and Signal Processing (ICASSP)}, pp. 9281--9285, 2022.

\bibitem{defossez2019demucsdeepextractormusic}
A.~Défossez, N.~Usunier, L.~Bottou, and F.~Bach, ``Demucs: Deep extractor for music sources with extra unlabeled data remixed,'' \emph{arXiv preprint arXiv:1909.01174}, 2019.

\bibitem{8707065}
Y.~Luo and N.~Mesgarani, ``{Conv-TasNet}: Surpassing ideal time–frequency magnitude masking for speech separation,'' \emph{IEEE/ACM Transactions on Audio, Speech, and Language Processing (TASLP)}, vol.~27, no.~8, pp. 1256--1266, 2019.

\bibitem{yang24h_interspeech}
H.~Yang, J.~Su, M.~Kim, and Z.~Jin, ``Genhancer: High-fidelity speech enhancement via generative modeling on discrete codec tokens,'' in \emph{{Interspeech}}, 2024, pp. 1170--1174.

\bibitem{wang2024selmspeechenhancementusing}
Z.~Wang, X.~Zhu, Z.~Zhang, Y.~Lv, N.~Jiang, G.~Zhao, and L.~Xie, ``{SELM}: Speech enhancement using discrete tokens and language models,'' in \emph{IEEE International Conference on Acoustics, Speech and Signal Processing (ICASSP)}, 2024, pp. 11\,561--11\,565.

\bibitem{kang2025llaseg1incentivizinggeneralizationcapability}
B.~Kang, X.~Zhu, Z.~Zhang, Z.~Ye, M.~Liu, Z.~Wang, Y.~Zhu, G.~Ma, J.~Chen, L.~Xiao, C.~Weng, W.~Xue, and L.~Xie, ``{LL}a{SE}-g1: Incentivizing generalization capability for {LL}a{MA}-based speech enhancement,'' in \emph{Annual Meeting of the Association for Computational Linguistics (Volume 1: Long Papers)}, 2025, pp. 13\,292--13\,305.

\bibitem{yao2025gensegenerativespeechenhancement}
J.~Yao, H.~Liu, C.~Chen, Y.~Hu, E.~Chng, and L.~Xie, ``{GenSE}: Generative speech enhancement via language models using hierarchical modeling,'' in \emph{International Conference on Learning Representations (ICLR)}, 2025.

\bibitem{sun2020emapproachnonautoregressiveconditional}
Z.~Sun and Y.~Yang, ``An {EM} approach to non-autoregressive conditional sequence generation,'' in \emph{International Conference on Machine Learning (ICML)}, 2020, pp. 9249--9258.

\bibitem{defossez2020realtimespeechenhancement}
A.~Défossez, G.~Synnaeve, and Y.~Adi, ``Real time speech enhancement in the waveform domain,'' in \emph{{Interspeech}}, 2020, pp. 3291--3295.

\bibitem{reddy2019scalablenoisyspeechdataset}
C.~K. Reddy, E.~Beyrami, J.~Pool, R.~Cutler, S.~Srinivasan, and J.~Gehrke, ``A scalable noisy speech dataset and online subjective test framework,'' in \emph{Interspeech}, 2019, pp. 1816--1820.

\bibitem{8031044}
C.~Karadagur Ananda~Reddy, N.~Shankar, G.~Shreedhar~Bhat, R.~Charan, and I.~Panahi, ``An individualized super-{G}aussian single microphone speech enhancement for hearing aid users with smartphone as an assistive device,'' \emph{IEEE Signal Processing Letters}, vol.~24, no.~11, pp. 1601--1605, 2017.

\bibitem{zorila2019investigationeffectivenessenhancementasr}
C.~Zorila, C.~Boeddeker, R.~S. Doddipatla, and R.~H{\"a}b-Umbach, ``An investigation into the effectiveness of enhancement in {ASR} training and test for {Chime-5} dinner party transcription,'' \emph{IEEE Automatic Speech Recognition and Understanding Workshop (ASRU)}, pp. 47--53, 2019.

\bibitem{Li_2025}
H.~Li, J.~Q. Yip, T.~Fan, and E.~S. Chng, ``Speech enhancement using continuous embeddings of neural audio codec,'' in \emph{IEEE International Conference on Acoustics, Speech and Signal Processing (ICASSP)}, Apr. 2025, pp. 1--5.

\bibitem{zeghidour2021soundstream}
N.~Zeghidour, A.~Luebs, A.~Omran, J.~Skoglund, and M.~Tagliasacchi, ``{SoundStream}: An end-to-end neural audio codec,'' \emph{IEEE/ACM Transactions on Audio, Speech, and Language Processing (TASLP)}, pp. 495--507, 2021.

\bibitem{xu2021anytimesamplingautoregressivemodels}
Y.~Xu, Y.~Song, S.~Garg, L.~Gong, R.~Shu, A.~Grover, and S.~Ermon, ``Anytime sampling for autoregressive models via ordered autoencoding,'' in \emph{International Conference on Learning Representations (ICLR)}, 2021.

\bibitem{rippel2014learningorderedrepresentationsnested}
O.~Rippel, M.~Gelbart, and R.~Adams, ``Learning ordered representations with nested dropout,'' in \emph{International Conference on Machine Learning (ICML)}, 2014, pp. 1746--1754.

\bibitem{guo2024socodecsemanticorderedmultistreamspeech}
H.~Guo, F.~Xie, K.~Xie, D.~Yang, D.~Guo, X.~Wu, and H.~Meng, ``{SoCodec}: A semantic-ordered multi-stream speech codec for efficient language model based text-to-speech synthesis,'' in \emph{IEEE Spoken Language Technology Workshop (SLT)}, 2024, pp. 645--651.

\bibitem{hsu2021hubertselfsupervisedspeechrepresentation}
W.-N. Hsu, Y.-H. Tsai, B.~Bolte, R.~Salakhutdinov, and A.~Mohamed, ``{HuBERT}: How much can a bad teacher benefit {ASR} pre-training?'' in \emph{IEEE International Conference on Acoustics, Speech and Signal Processing (ICASSP)}, 2021, pp. 6533--6537.

\bibitem{zhang2024speechtokenizerunifiedspeechtokenizer}
X.~Zhang, D.~Zhang, S.~Li, Y.~Zhou, and X.~Qiu, ``{SpeechTokenizer}: Unified speech tokenizer for speech language models,'' in \emph{International Conference on Learning Representations (ICLR)}, 2024.

\bibitem{ren2022revisitingoversmoothnesstextspeech}
Y.~Ren, X.~Tan, T.~Qin, Z.~Zhao, and T.-Y. Liu, ``Revisiting over-smoothness in text to speech,'' in \emph{Annual Meeting of the Association for Computational Linguistics (Volume 1: Long Papers)}, May 2022, pp. 8197--8213.

\bibitem{ma2018noisecontrastiveestimationnegative}
Z.~Ma and M.~Collins, ``Noise contrastive estimation and negative sampling for conditional models: Consistency and statistical efficiency,'' in \emph{Conference on Empirical Methods in Natural Language Processing (EMNLP)}, 2018, pp. 3698--3707.

\bibitem{kumar2023highfidelityaudiocompressionimproved}
R.~Kumar, P.~Seetharaman, A.~Luebs, I.~Kumar, and K.~Kumar, ``High-fidelity audio compression with improved {RVQGAN},'' in \emph{International Conference on Neural Information Processing Systems (NeurIPS)}, 2023.

\bibitem{li2024masksrmaskedlanguagemodel}
\BIBentryALTinterwordspacing
X.~Li, Q.~Wang, and X.~Liu, ``Masksr: Masked language model for full-band speech restoration,'' 2024. [Online]. Available: \url{https://arxiv.org/abs/2406.02092}
\BIBentrySTDinterwordspacing

\bibitem{7178964}
V.~Panayotov, G.~Chen, D.~Povey, and S.~Khudanpur, ``Librispeech: An {ASR} corpus based on public domain audio books,'' in \emph{IEEE International Conference on Acoustics, Speech and Signal Processing (ICASSP)}, 2015, pp. 5206--5210.

\bibitem{reddy2020interspeech}
C.~K. Reddy, V.~Gopal, R.~Cutler, E.~Beyrami, R.~Cheng, H.~Dubey, S.~Matusevych, R.~Aichner, A.~Aazami, S.~Braun \emph{et~al.}, ``The {Interspeech} 2020 deep noise suppression challenge: Datasets, subjective testing framework, and challenge results,'' in \emph{Interspeech}, 2020.

\bibitem{thiemann2013demand}
J.~Thiemann, N.~Ito, and E.~Vincent, ``The diverse environments multi-channel acoustic noise database ({DEMAND}): A database of multichannel environmental noise recordings,'' \emph{Proceedings of Meetings on Acoustics}, vol.~19, no.~1, 2013.

\bibitem{reddy2021dnsmosnonintrusiveperceptualobjective}
C.~K. Reddy, V.~Gopal, and R.~Cutler, ``{DNSMOS}: A non-intrusive perceptual objective speech quality metric to evaluate noise suppressors,'' in \emph{IEEE International Conference on Acoustics, Speech and Signal Processing (ICASSP)}, 2021, pp. 6493--6497.

\bibitem{libera2025focalcodec}
L.~{Della Libera}, F.~Paissan, C.~Subakan, and M.~Ravanelli, ``{FocalCodec}: Low-bitrate speech coding via focal modulation networks,'' in \emph{International Conference on Neural Information Processing Systems (NeurIPS)}, 2025.

\bibitem{dellalibera2025focalcodecstreamstreaminglowbitratespeech}
L.~{Della Libera}, C.~Subakan, and M.~Ravanelli, ``{FocalCodec-Stream}: Streaming low-bitrate speech coding via causal distillation,'' \emph{arXiv preprint arXiv:2509.16195}, 2025.

\bibitem{Liu_2022}
H.~Liu, X.~Liu, Q.~Kong, Q.~Tian, Y.~Zhao, D.~Wang, C.~Huang, and Y.~Wang, ``Voicefixer: A unified framework for high-fidelity speech restoration,'' in \emph{Proc. Interspeech}, Sep. 2022, pp. 4232--4236.

\bibitem{Zhang_2025}
J.~Zhang, J.~Yang, Z.~Fang, Y.~Wang, Z.~Zhang, Z.~Wang, F.~Fan, and Z.~Wu, ``Anyenhance: A unified generative model with prompt-guidance and self-critic for voice enhancement,'' \emph{IEEE Transactions on Audio, Speech and Language Processing}, vol.~33, pp. 3085--3098, 2025.

\end{thebibliography}
\endgroup

\vspace{12pt}

\end{document}